\documentclass[aps,prb,preprint,superscriptaddress,showpacs]{revtex4}
\usepackage{graphicx}
\usepackage{dcolumn}
\usepackage{multirow}
\usepackage{subfigure}
\usepackage[usenames]{color}

\begin{document}

\title{High order vibration modes of glass embedded AgAu nanoparticles}

\author{S. Adichtchev}
\author{S. Sirotkin}
\affiliation{Laboratoire de Physico-Chimie des Mat\'eriaux Luminescents, Universit\'e de Lyon, Universit\'e Claude Bernard Lyon 1, UMR 5620 CNRS, 69622 Villeurbanne, France}
\author{G. Bachelier}
\affiliation{Laboratoire de Spectroscopie Ionique et Mol\'eculaire, Universit\'e de Lyon, Universit\'e Claude Bernard Lyon 1, UMR 5579 CNRS, 69622 Villeurbanne, France}
\author{L. Saviot}
\affiliation{Institut Carnot de Bourgogne, UMR 5209 CNRS-Universit\'e de Bourgogne, 9 av. A. Savary, BP 47870, F-21078 DIJON Cedex, France}
\author{S. Etienne}
\affiliation{Ecole Europ\'eenne d'Ing\'enieurs en G\'enie des Mat\'eriaux, 6 rue Bastien Lepage, 54010 Nancy Cedex, Nancy Universit\'e, France}
\author{B. Stephanidis}
\author{E. Duval}
\author{A. Mermet}
\affiliation{Laboratoire de Physico-Chimie des Mat\'eriaux Luminescents, Universit\'e de Lyon, Universit\'e Claude Bernard Lyon 1, UMR 5620 CNRS, 69622 Villeurbanne, France}

\date{\today}

\begin{abstract}
High resolution low frequency Raman scattering measurements from embedded AgAu nanoparticles unveil efficient scattering by harmonics of both the quadrupolar and the spherical modes. Comparing the experimental data with theoretical calculations that account for both the embedding medium and the resonant Raman process enables a very complete description of the observed multiple components in terms of harmonics of both the quadrupolar and spherical modes, with a dominating Raman response from the former ones. It is found that only selected harmonics of the quadrupolar mode contribute significantly to the Raman spectra in agreement with earlier theoretical predictions.
\end{abstract}

\pacs{63.22.-m,78.30.-j,78.67.-n}
\maketitle

Over the past twenty years, low frequency vibrational modes from solid nanoparticles have emerged as original features of nanometer sized matter. These modes arise from the recombination of acoustic modes that turn from {\it extended} to {\it confined} upon reducing the size of the solid.
Raman scattering is a dedicated technique to probe nanoparticle oscillation modes provided the used spectrometers enable to reach the low frequency range where these modes typically show up ($\sigma < 10$ cm$^{-1}$). It has proved to be a powerful tool for a quick and non-destructive evaluation of the sizes of the nanoparticles, especially in the case of metallic ones for which Raman scattering is performed in resonance with the localized surface plasmon excitations. Under these conditions, the Raman activity of selected vibration modes of the nanoparticles is strongly enhanced so that intense signals can be obtained from very diluted nanoparticle assemblies\cite{Courty2002,Stephanidis2007}. Metallic nanoparticles have allowed significant progress in the understanding of nanoparticle modes like shape anisotropy effects from Ag nanocolumns and nanolentils\cite{Margueritat2007} and elastic anisotropy effects in Au nanocrystals \cite{Stephanidis2007,Portales2008}. Concomitantly, the theoretical description of the Raman spectra from nanoparticle vibrations has significantly improved over the years: while the free sphere model of H. Lamb\cite{Lamb1881}, combined with appropriate Raman selection rules\cite{Duval1992}, initially provided satisfactory predictions of the mode frequencies\cite{Duval1986}, coupling with the
vibrations inside a surrounding matrix in the case of \textsl{embedded}
nanoparticles have been well accounted for\cite{Dubrovskiy1981,Verma1999}. On the other hand, quantum calculations on \textsl{free} metal nanospheres, taking into account the vibration-plasmon coupling, have shown to be able to describe the resonant Raman spectrum in a rather complete way both in terms of frequencies and Raman intensities\cite{Bachelier2004}. In particular, this latter work predicted that the Raman intensities from the quadrupolar mode harmonics are not a monotonic function of the mode index. The measurements presented in this work, which show an unusually high number of low-frequency Raman peaks, are an experimental proof of the non monotonic variation of intensity. Thanks to the observation of high order harmonics, we show that the interpretation of low frequency Raman data from metallic nanoparticles requires taking into account the coupling of the surface plasmon with the different vibrations but also the influence of the surrounding
matrix as well as the elastic anisotropy inside the nanoparticle. 

\begin{figure}
 \includegraphics[width=\columnwidth]{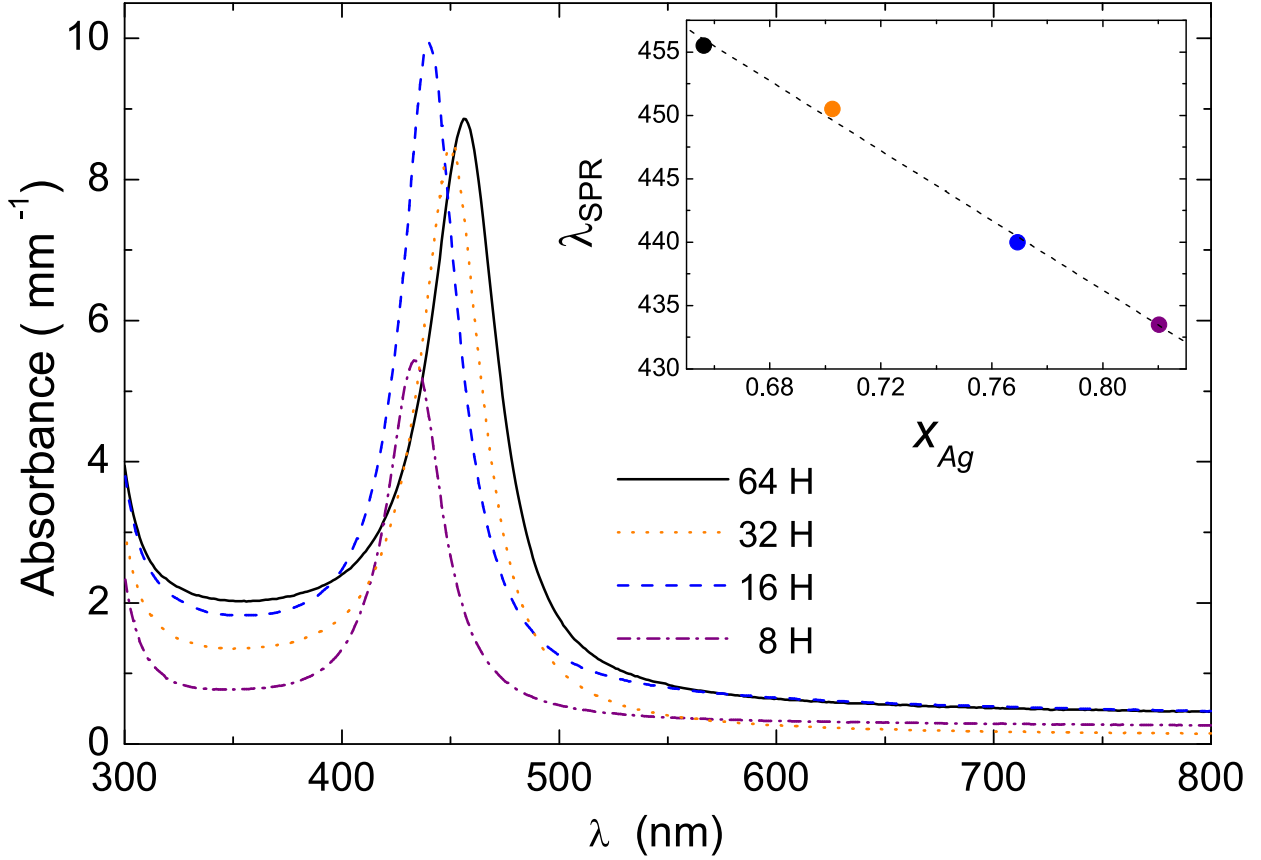}
 \caption{\label{absorption} (color online) Absorption spectra of glasses containing Ag$_x$Au$_{1-x}$ nanoparticles for different annealing times. The inset shows the SPR peak position as a function of the Ag composition, the dashed line is a guide to the eye.}
\end{figure}

Bimetallic AgAu nanoparticles were produced in a bulk multicomponent glass, starting from the incorporation of silver and gold salts to the initial oxides powder mix. After the colorless glass is made, prolonged annealings slightly above the glass transition temperature generates a characteristic amber shade that arises from the surface plasmon resonance (SPR) of the nucleated metallic nanoparticles. The optical characterization of four samples annealed at 490$^\circ$C over 8 H, 16 H, 32 H and 64 H shows a single surface plasmon resonance peak around 445 nm (Figure~\ref{absorption}). 
This value typically lies between that of pure silver (414 nm) and that of pure gold (529 nm) when embedded as pure nanoparticles with comparable sizes in the same glass, using the same synthesis route. Upon increasing the annealing time, the AgAu SPR is red shifted thus reflecting a gold enrichment of the bimetallic nanoparticles. Following a classical interpretation of these data \cite{Link1999}, one can derive the composition ratio $x$ of the so-produced Ag$_x$Au$_{1-x}$ nanoparticles as a function of annealing time $t_a$ using a linear combination of the SPR peak positions :
\begin{equation}
\lambda_{Ag_xAu_{1-x}}(t_a)=x\lambda_{Ag}(t_a)+(1-x)\lambda_{Au}(t_a)
\end{equation}
where $\lambda_{Ag}(t_a)$ and $\lambda_{Au}(t_a)$ are respectively the pure Ag and the pure Au nanoparticles produced in the same glass, at exactly the same annealing times. The so-derived values of $x$ are found to decrease from 0.82 to 0.66 as the annealing times respectively increase from 8~H to 64 H.

The low frequency Raman scattering spectra were recorded with a six pass tandem Fabry-Perot interferometer using the 532 nm line of a continuous doubled Yttrium Aluminum Garnet laser of 150 mW power. The scattered light was collected in backscattering geometry and the free spectral range set to 550 GHz. The finesse and optical contrast of such interferometer are typically greater than 100 and $10^{10}$ respectively. More details on the tandem Fabry-Perot interferometer can be found elsewhere\cite{Sandercock1982, Mock1987}. 

\begin{figure}
 \includegraphics[width=\columnwidth]{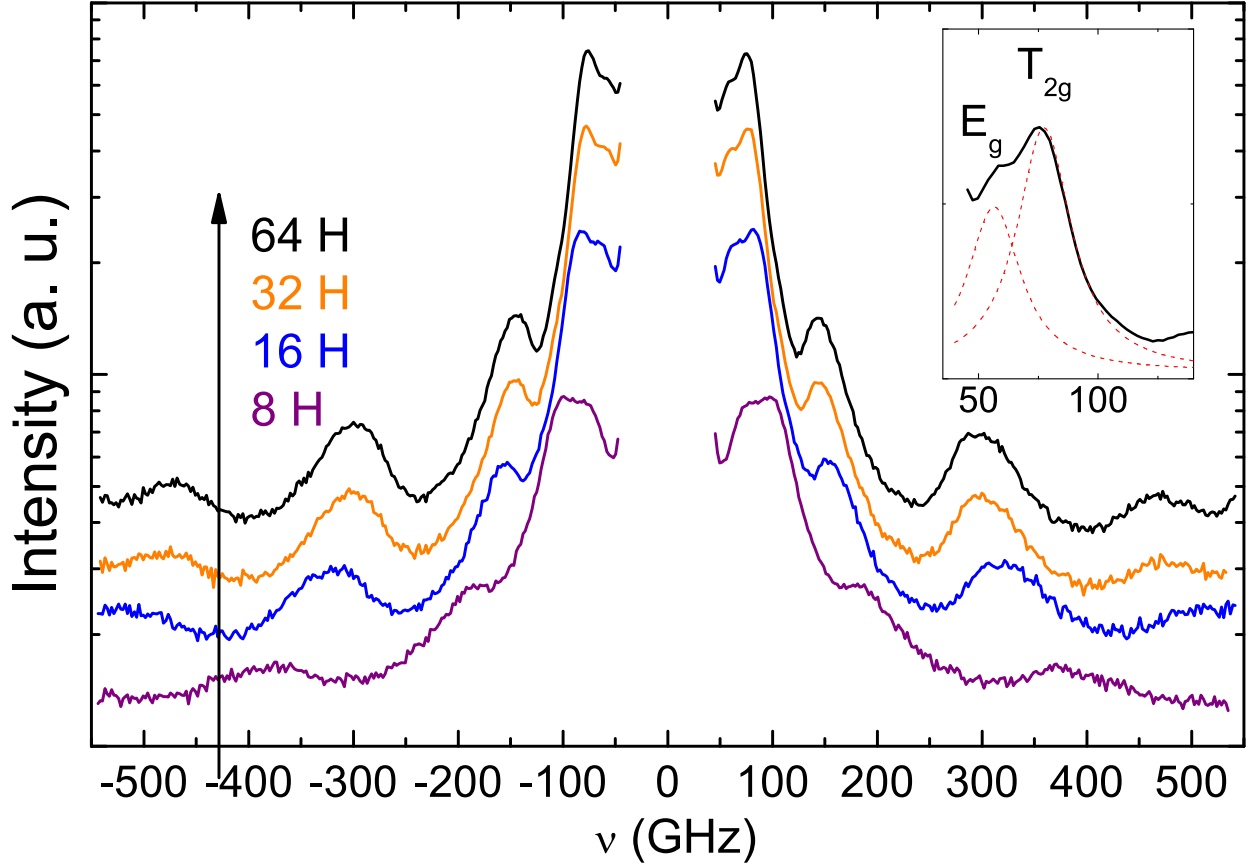} 
 \caption{\label{AllRaman} (color online) Polarized (VV) low frequency Raman spectra (in Log-linear scale) of AgAu nanoparticles embedded in glasses annealed over different times. Inset: enlargement of the lower ($E_g$,$T_{2g}$) doublet for the 64 H sample.}
\end{figure}
As a general illustration, Figure \ref{AllRaman} reports the polarized spectra (with incident and scattered light polarized both vertically, VV) of the afore characterized samples.
Quite similarly to recent observations\cite{Stephanidis2007}, these spectra are remarkably richer than those usually reported for metallic nanoparticles in the sense that they evidence more components than the usually two observed ones, \textit{i.e.} the fundamental vibrations of both the quadrupolar mode $(\ell=2,n=1)$ and the spherical mode $(\ell=0, n=1)$ ($\ell$ denotes the angular momentum and $n$ the harmonic rank). Four main bands can already be identified, some of which featuring subcomponents. For purposes of illustration, we will consider in the rest of the paper only the longest annealed sample (Ag$_{0.66}$Au$_{0.34}$), provided that the same analysis apply equally well to all samples.

A classical way of identifying the several components of a low frequency Raman scattering spectrum is to compare the polarized (VV) and depolarized (VH) spectra since quadrupolar modes and spherical modes behave differently with regards to polarization.
In a first step, we focus on the VH spectrum which is more simple as it only features contributions from the depolarized quadrupolar modes. A multi-Lorentzian decomposition of the VH spectrum reveals 5 components, two of them forming an intense lower frequency doublet (Figure~\ref{AllRaman} Inset) evidenced thanks to the high resolution Raman setup. Following recent works\cite{Stephanidis2007,Portales2008}, such doublet can be safely identified with the splitting of the fundamental mode of the quadrupolar vibration ($\ell=2, n=1$) due to the high crystallinity of the nanoparticles (\textit{i.e.} elastic anisotropy). Indeed, the observed frequency ratio between the lower and the larger components of the doublet, respectively labeled as $E_g$ and $T_{2g}$ according to the irreducible representation of the corresponding vibrations of a face centered cubic nano\textit{crystal}, is close to the value expected for free Au or free Ag nanocrystals\cite{Portales2008} ($\frac{\nu_{T_{2g}}}{\nu_{E_g}}\simeq1.4$ \textit{vs} $\sim 1.6$ respectively).

\begin{figure}
 \includegraphics[width=\columnwidth]{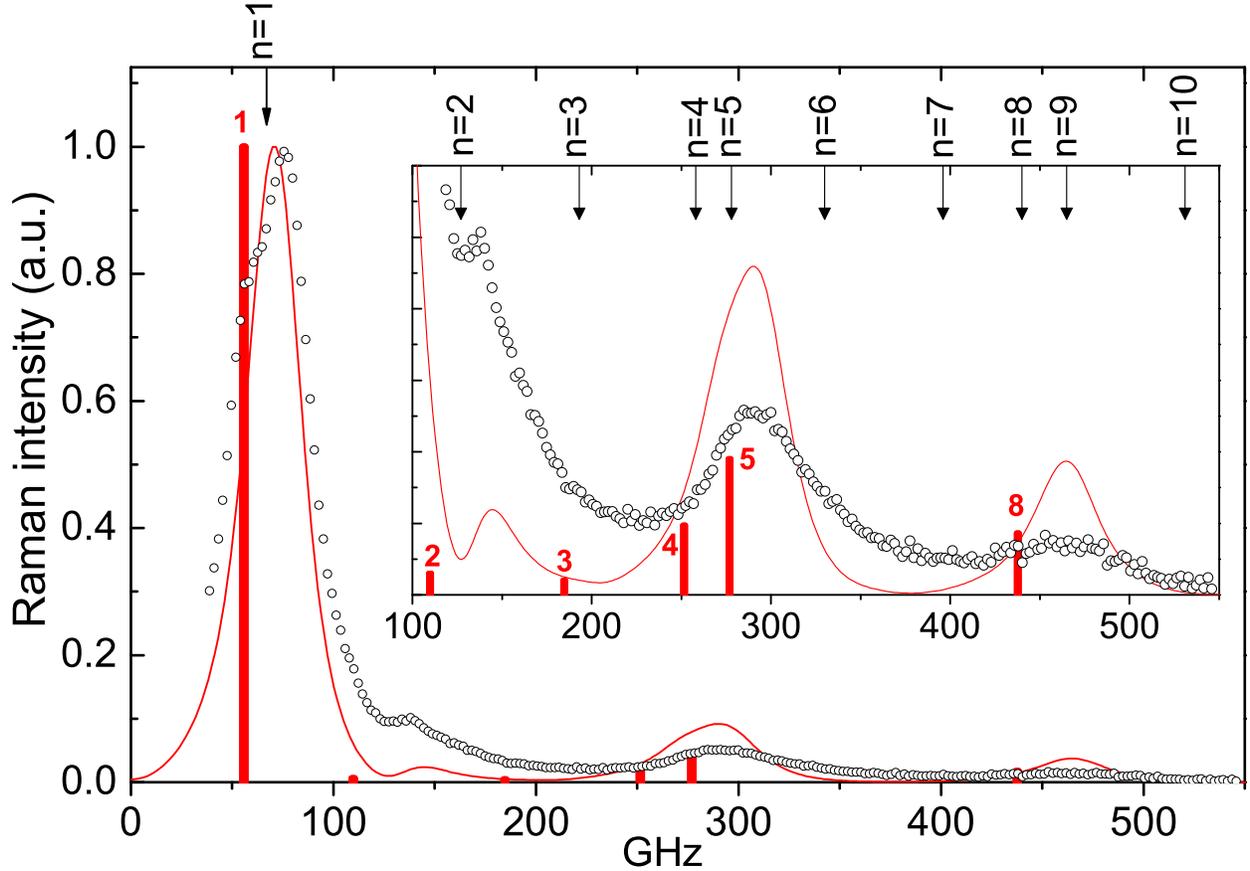} 
 \caption{\label{GB_VH} (color online) Experimental (symbols) and calculated (solid line) depolarized (VH) Raman spectra of the Ag$_{0.66}$Au$_{0.34}$ sample with a diameter of 23.5 nm (the spectra were normalized with respect to the $(\ell=2,n=1)$ peak); high frequency enlargement in the inset. The thick bars are the Raman intensities calculated for the same nanoparticle considered as \textit{free}, labeled according to their $n$ index (only harmonics with non negligible intensities are shown). Top scale arrows: CFM predictions for the $\ell=2$ mode with the same \textit{embedded} nanoparticle.}
\end{figure}

As suggested by previous theoretical
works\cite{Bachelier2004,Saviot2005}, the higher frequency components of
the depolarized Raman spectrum (approximately located at 150, 300 and
460 GHz) may be ascribed to selected high order harmonics of the
quadrupolar mode.
Similar predictions have been made recently for CdS$_x$Se$_{1-x}$
nanoparticles using different Raman scattering
mechanisms\cite{Ristic2008}.
In the following we compare our experimental results
to Raman Scattering Intensity (RSI) calculations based on the approach
developed by Bachelier \textit{et al}\cite{Bachelier2004}. In this
approach, the low frequency Raman spectrum of a metallic nanoparticle is
generated from the calculation of the Raman transition probabilities
using a three step scattering process involving the dipolar plasmon
state. For the herein investigated case, this calculation method was
extended to the case of an \textit{embedded} nanoparticle, through the
determination of the continuum frequency spectrum of the
nanoparticle-matrix system. A perfect contact between the particle and
the matrix was assumed in order to apply the standard boundary
conditions between the two media. The full derivation of the vibration
modes and their normalization is out of the scope of the present paper
and will be described in details elsewhere. In addition to the
excitation wavelength, the input parameters of these calculations are
the matrix acoustic parameters (density $\rho=2.97$~g.cm$^{-3}$;
$V_L=5020$~m.s$^{-1}$ and $V_T=3010$~m.s$^{-1}$, as derived from
Brillouin light scattering measurements on the samples) together with
those of the bimetallic nanoparticles that were assumed to be alloys and
thus obeying linear relationships as a function of $x$.\cite{Linear}
Figure~\ref{GB_VH} compares the experimental spectrum of the
Ag$_{0.66}$Au$_{0.34}$ sample to its calculated counterpart. The overall
aspect of the Raman spectrum is rather well reproduced in terms of band
positions and relative intensities. For further discussion, the top
scale of Figure~\ref{GB_VH} indicates the predictions of the Complex
Frequency Method\cite{Dubrovskiy1981,Verma1999} (CFM), which enables a rapid evaluation of the band positions for embedded nanoparticles yet without yielding the Raman intensities. In order to identify the filiation of the Raman bands with harmonics of the quadrupolar vibration from a single nanoparticle, Figure~\ref{GB_VH} shows RSI spectra for the same nanoparticle but considered as \textit{free}. In the absence of sufficiently reliable Transmission Electron Microscopy data due to the buried nature of the nanoparticles, the mean diameter of the nanoparticles is derived from the best fit of the lower frequency doublet with a single average ($\ell=2,n=1$) line since at this stage the RSI calculations are unable to account for the elastic anisotropy splitting (nanocrystals are assumed as isotropic nanoparticles). The derived diameter for this system is 23.5 nm, as confirmed from the analysis of the polarized spectrum that shows in addition the spherical mode contributions.

\begin{figure}
 \includegraphics[width=\columnwidth]{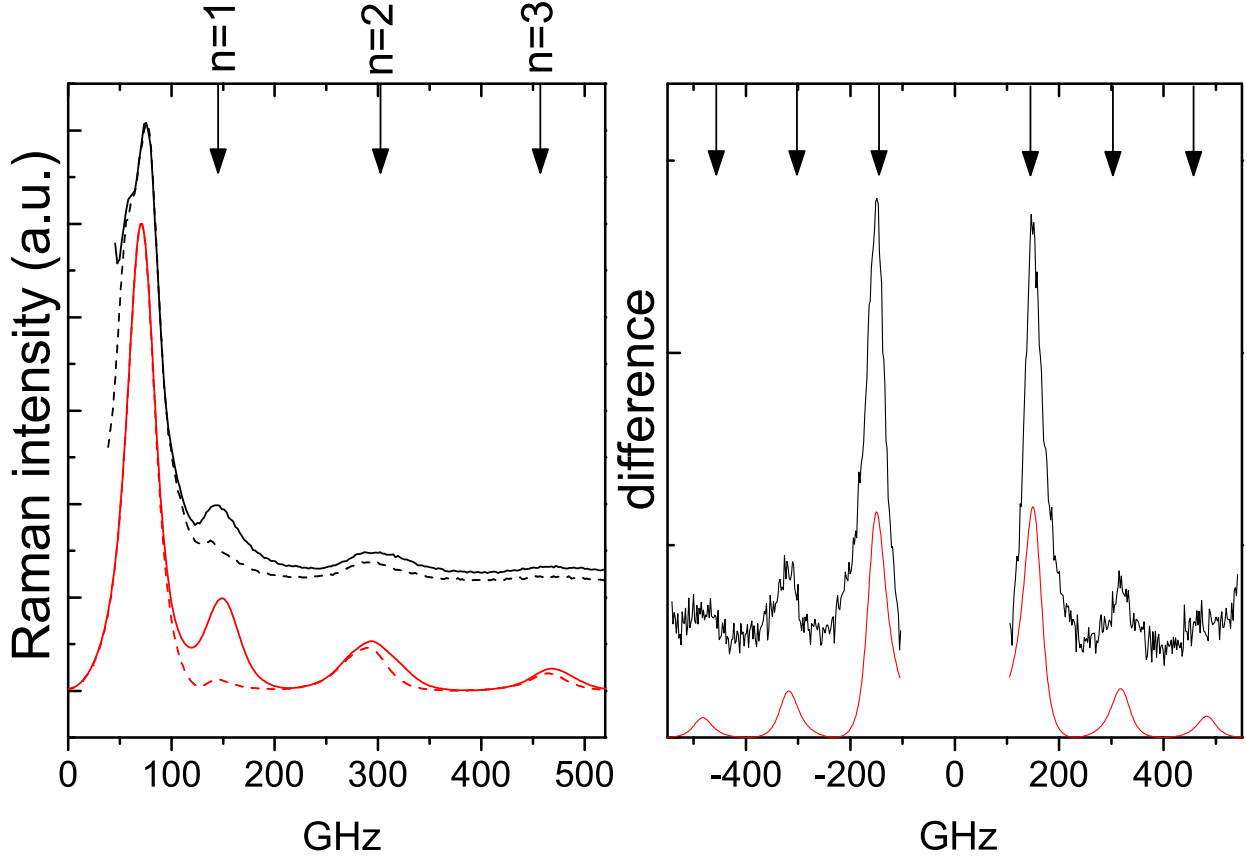} 
 \caption{\label{GB_VV} (color online) \textit{Left} : Experimental (upper set) and calculated (lower set) of the VV (solid) and VH (dashed) Raman profiles of the Ag$_{0.66}$Au$_{0.34}$ sample (intensities were normalized with respect to the ($\ell=2, n=1$) peak). \textit{Right}: Raman difference spectrum ($I_{VV}-I_{VH}$), experimental (thick line) and RSI calculated (thin line). Top scales: CFM predictions for the $\ell=0$ mode.}
\end{figure}

Figure~\ref{GB_VV} compares the VV and VH experimental spectra of the Ag$_{0.66}$Au$_{0.34}$ sample, together with those derived from the RSI calculations. These comparisons show that the spherical mode contributions are rather small and they significantly overlap with those of the quadrupolar mode. They are better evidenced by the Raman difference spectrum $I_{VV}-k.I_{VH}$, where $k$ is a scaling factor chosen such that the $(\ell=2, n=1)$ peaks are normalized (Fig.~\ref{GB_VV} left). The positions derived from the difference are found to well agree with the CFM predictions for the $(\ell=0, n=1,2,3)$ modes, using the same nanoparticle size as that derived from the quadrupolar mode analysis; they are equally well consistent with the RSI predictions of band profile changes induced by the change of polarization.

The present observation of an usually rich low frequency Raman spectrum from embedded metallic nanoparticles, and its comparison with RSI and CFM predictions, allows to capture several ingredients that are essential for a correct interpretation of similar data.
The proper identification of the several Raman bands requires, in a first stage, to compare their behaviour as a function of polarization as they may result from the overlap of several contributions (Fig.~\ref{GB_VV}). The detailed analysis of the depolarized spectrum, \textit{i.e.} of the quadrupolar mode contributions, shows several important points. The first one is that introducing the embedding medium affects the position of the bands as displayed by the frequency mismatch of the observed bands (symbols in Fig.~\ref{GB_VH}) or the CFM positions (top arrows) with the frequencies calculated in the free case (thick bars). This effect, which is essentially due to the relatively large mass density of the used glass, is most severe on the $(\ell=2,n=1)$ peak, which is often used for size evaluation due to its prominence. Quantitatively, using a free case estimation based on the position of the $(\ell=2,n=1)$ band leads to a $19\%$ smaller nanoparticle diameter.
The second result is that from the frequency position matching between the experimental curve and the calculated RSI and CFM contributions, it comes out that the high frequency components are affiliated with selected harmonics, or at least harmonic groups, of the single nanoparticle vibration: the bands near 290 GHz and 460 GHz clearly identify with respectively the $n=4,5$ and $n=8$ harmonic components of the quadrupolar mode. Their respective Raman cross-sections are relatively consistent with those predicted in the free sphere case (delta functions in Fig.~\ref{GB_VH})\cite{Bachelier2004}. The situation is somewhat less clearcut with the weak band at 150 Ghz which well identifies with the $n=2$ harmonic in terms of frequency while its intensity is considerably stronger than that expected from the free sphere evaluation: according to RSI calculations, it is enhanced by a factor 4 going from the free case to the embedded case while the other harmonics are enhanced by a factor 2 (its stronger intensity in the experimental spectrum is certainly due to its merging with the foot of the $T_{2g}$ component that the RSI calculations are unable to account for at the present stage). This enhancement is rooted in the coupling with the matrix that significantly increases the displacement at the surface of the nanoparticle, for this particular harmonic, making it a potential good probe of the coupling interface. Another possible interpretation is that the 150 GHz band results from the redistribution of mode frequencies that arises from the anistropy effect, as is observed from the splitting of the $(\ell=2, n=1)$ component. These two interpretations differ from that given for a similar band observed from similar Au doped glasses\cite{Stephanidis2007}.

Finally, the comparison of the RSI and CFM calculations with the experimental data provides a further interesting aspect related to the specificity of \textit{resonant} Raman scattering for metallic nanoparticles. Indeed, one observes that while the frequency positions of the spherical mode bands agree well with those predicted by either RSI and CFM calculations (Fig.~\ref{GB_VV}), the CFM predictions for the quadrupolar mode harmonics are systematically slightly red shifted with respect to either the experimental or the RSI calculated bands (Fig.~\ref{GB_VH}). This is due to the dual nature of the quadrupolar modes (longitudinal vs transverse as defined in \cite{Saviot2005}) which is also present inside a given CFM resonance. The plasmon-vibration coupling is therefore significant only for a subset of modes inside the frequency range defined by a CFM peak which can lead to a shift of the observed maxima. This situation does not occur for the spherical mode. As a result, the CFM and RSI peak positions match and these modes can be more reliably used for the size evaluation of the nanoparticles provided they are safely extracted from the dominating quadrupolar mode response.

The first observation of high order quadrupolar and spherical harmonic modes from AgAu nanoparticles embedded in a glass has allowed us to rationalize the interpretation of low frequency Raman spectra from metallic nanoparticles. In accord with theoretical predictions, one observes that Raman scattering by the quadrupolar modes is not a monotonic decreasing function of the harmonic index so that only selected harmonics participate to the Raman spectrum with differing efficiencies. Due to stronger coupling with the plasmon excitations, the Raman spectrum is strongly dominated by high order harmonics of the quadrupolar modes that overlap with the weaker polarized signatures from the spherical modes. In such case, a detailed polarization analysis of the Raman bands is therefore essential to check the band assignments that are uniquely based on frequency position matchings\cite{Portales2001a,Nelet2004}.

\begin{acknowledgments}
The authors acknowledge N. Del Fatti and F. Vall\'ee for their relevant comments.
\end{acknowledgments}

\bibliography{AuAg3}

\end{document}